\documentclass[journal]{IEEEtran}

\usepackage{graphicx}
\usepackage{amsmath}
\usepackage{amssymb}
\usepackage{comment}
\usepackage{float}
\usepackage[dvipsnames]{xcolor}
\usepackage{ifpdf}
\usepackage{cite}
\usepackage{hyperref}
\usepackage{enumitem}
\usepackage{algorithm2e}

\makeatletter

\renewcommand{\@algocf@capt@plain}{above}
\makeatother

\begin{document}
\title{Mitigating the Risk of Voltage Collapse using\\Statistical Measures from PMU Data}

\author{Samuel Chevalier,~\IEEEmembership{Student Member,~IEEE,}
        Paul D.~H.~Hines,~\IEEEmembership{Senior Member,~IEEE}
\thanks{S.~Chevalier is with the Department of Mechanical Engineering at the Massachusetts Institute of Technology, Cambrdige, MA 02139 (e-mail: schev@mit.edu).

P.D.H.~Hines is with the Dept.~of Electrical and Biomedical Engineering at the University of Vermont, Burlington, VT 05405 USA (e-mail: paul.hines@uvm.edu).}}

\maketitle

\begin{abstract}
With the continued deployment of synchronized Phasor Measurement Units (PMUs), high sample rate data are rapidly increasing the real time observability of power systems. Prior research has shown that the statistics of these data can provide useful information regarding network stability, but it is not yet known how this statistical information can be actionably used to improve power system stability.
To address this issue, this paper presents a method that gauges and improves the voltage stability of a system using the statistics present in PMU data streams. Leveraging an analytical solver to determine a range of ``critical'' bus voltage variances, the presented methods monitor raw statistical data in an observable load pocket to determine when control actions are needed to mitigate the risk of voltage collapse. A simple reactive power controller is then implemented, which acts dynamically to maintain an acceptable voltage stability margin within the system. Time domain simulations on 3-bus and 39-bus test cases demonstrate that the resulting statistical controller can out-perform more conventional feedback control systems by maintaining voltage stability margins while loads simultaneously increase and fluctuate.
\end{abstract}


\IEEEpeerreviewmaketitle

\section{Introduction}\label{introduction}

\IEEEPARstart{I}{n} order to optimize the use of limited infrastructure, 
transmission systems frequently operate close to their stability or security limits.
Although economically advantageous~\cite{Dobson:2007}, this can lead to elevated blackout risk given the fluctuating nature of supply and demand~\cite{Ohno:2006}. Consequently, stability margin estimation is an essential tool for power system operators. Predicting the onset of voltage instability, though, is often made difficult by reactive support systems and tap changing transformers that hold voltage magnitudes high as load increases.
Although voltage support is essential for reliable operations, these controls can sometimes hide the fact that a system is approaching a voltage stability limit, particularly when operators and control systems rely on voltage magnitude data for control decisions.

Across a variety of complex systems, there is increasing evidence that indicators of emerging critical transitions 
can be found in the statistics of state variable time series data~\cite{Scheffer:2009}. Termed Critical Slowing Down (CSD)~\cite{Wissel:1984}, this phenomenon most clearly appears 
as elevated variance and autocorrelation in time-series data~\cite{Dakos:2012}. 
More recently, CSD has been successfully investigated in the power systems literature, and strong connections have been drawn between bifurcation theory and the elevation of certain statistics in voltage and current time series data~\cite{Ghanavati:2014,Podolsky:2013,Cotilla:2012,Ghanavati:2015}. In particular,~\cite{Ghanavati:2015} presents a method for analytically calculating the time series statistics associated with a stochastically forced dynamic power system model. These results are leveraged in this paper in order to predict key statistics of a power system that is approaching a critical transition. Others, such as~\cite{Sodhi:2012}, have developed control methods that use voltage magnitude declination rate measurements, but do not explicitly use statistical information as are presented in this paper. 

As reviewed in~\cite{Lerm:2003}, power systems are liable to experience a variety of critical transitions, including Hopf, pitchfork, and limit-induced bifurcations. 
This paper is
primarily concerned with the slow load build up, reactive power shortages, and other Long Term Voltage Stability (LTVS) processes that may contribute to a Saddle-Node bifurcation of the algebraic power flow equations. Classic voltage stability, which refers to a power system's ability to uphold steady voltage magnitudes at all network buses after experiencing a disturbance, is lost after a network experiences this sort of bifurcation~\cite{Kundur:2004}. The methods in this paper aim to preserve such voltage stability and thus prevent a system from experiencing voltage collapse.

The goal of this paper is to describe and evaluate a control system that uses the variance of bus voltages to reduce the probability of voltage collapse in a stochastic power system. 
This control system leverages a number of innovative tools to perform this task. 
The first uses the load scaling factor from the Holomorphic Embedding Load Flow Method (HELM)~\cite{Trias:2012} to represent a slowly varying stochastic variable, such as changes in overall load levels over time. 
Second, a First Passage Process (FPP)~\cite{Redner:2001} is used to identify critical loading thresholds given the statistics of the slower load changes. Finally, a full order dynamical system model is used to analytically predict the expected algebraic variable covariance matrix of the system, given stochastic load noise excitation, for the previously identified critical loading level. The associated critical variances from this matrix are then used as a reference signal to control a static VAR compensator (SVC). This reference signal is dynamically updated as network configurations change and equilibriums shift. In building this controller (see Fig.~\ref{fig: VBC_Model}), this paper combines static algebraic voltage collapse analysis through HELM, the first passage probability, statistical estimation (based on system model and load noise assumptions) and network feedback in order to leverage the statistical properties of PMU data to reduce the risk of voltage instability.

The remainder of this paper is organized as follows. Section~\ref{Background} motivates the use of voltage variance as an indicator of stability and presents the methods
employed in the statistical controller which is developed in this paper. Section~\ref{Controller Outline} describes the new statistical controller along with two conventional reactive power controllers that are used to benchmark the results against. Section~\ref{Test Results} demonstrates the statistical controller on a 3 bus power system consisting of a generator, a load bus, and an SVC bus. For further validation, Section~\ref{Test Results 39} tests the controllers on the IEEE 39 bus system. Finally, conclusions and  ideas for future research are presented in Section~\ref{Conclusion}.

\section{Background}\label{Background}
This section motivates the 
use of bus voltage variance as a measure of stability and presents the methods and tools 
used to build our statistical controller. 

\subsection{Bus Voltage Variance in a 2 Bus Power System}
A variety of systems, such as capacitor banks, tap changing transformers, and various Flexible AC Transmission System (FACTS) devices, 
are employed in power systems to ensure that voltage magnitudes remain sufficiently high.
As a result, voltage magnitudes are an imperfect indicator of the proximity of a system to its voltage stability limits.
To understand how an overloaded system with high voltage magnitudes may have a compromised voltage stability margin, the definition of loading margin in~\cite{Greene:1997} (p.262) is first given: ``For a particular operating point, the amount of additional load in a specific pattern of load increase that would cause a voltage collapse is called the loading margin.''

\begin{figure}[H]
\noindent \begin{centering}
\includegraphics[scale=0.95]{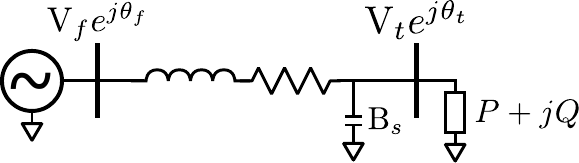}
\par\end{centering}
\caption{\label{fig: 2_Bus_Shunt}Two bus model with generator, load and shunt capacitor $\rm B_s$}
\vspace{0.2in}
\noindent \begin{centering}
\includegraphics[scale=0.45]{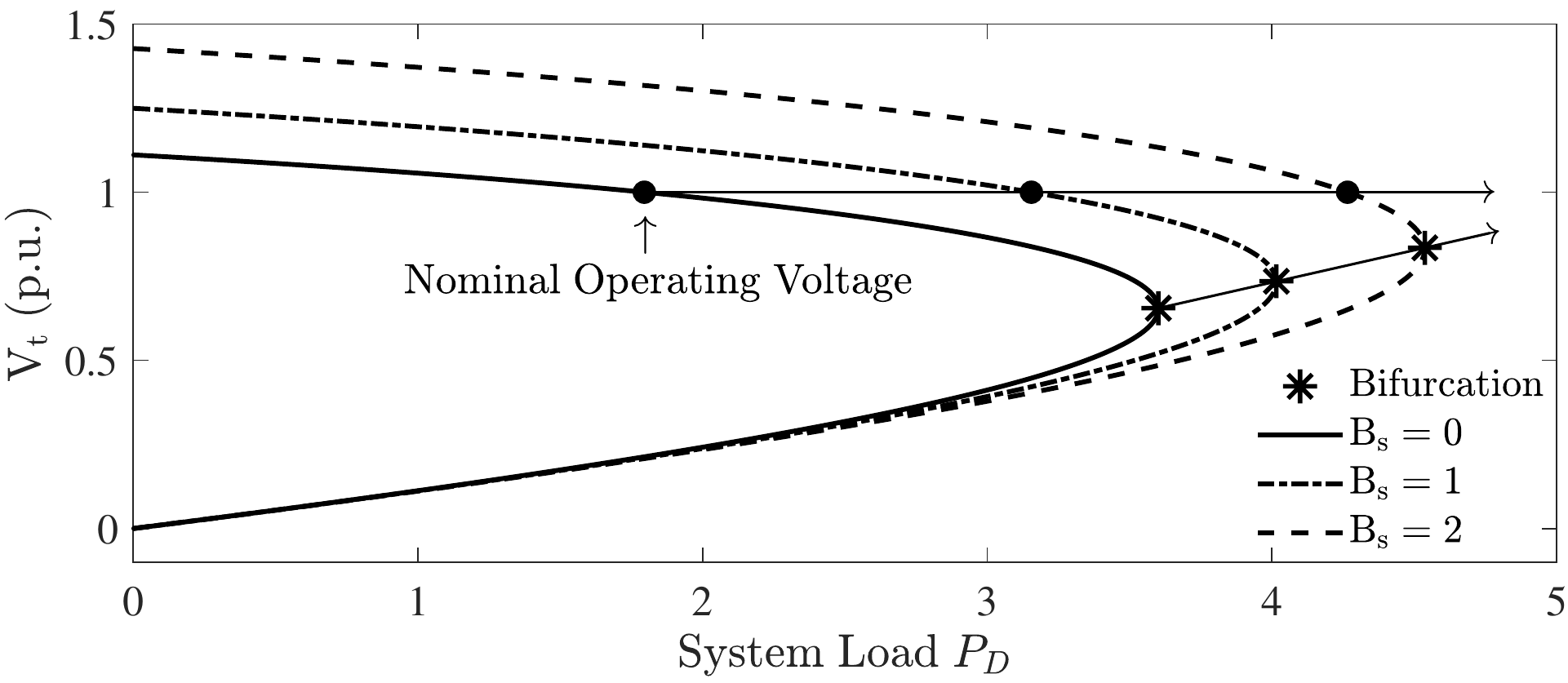}
\par\end{centering}
\protect\caption{\textcolor{black}{\label{fig: Nose_Curves}Load bus voltage as a function of $P_{D}$ for various shunt values. As more reactive power is injected into the system, the voltage magnitude at the point of bifurcation drifts upward toward nominal operating voltage.}}
\end{figure}

Consider the system in Fig.~\ref{fig: 2_Bus_Shunt}, where capacitive shunt support ${\rm B}_s$ and a constant power load $P+jQ$ are placed at the ``to" bus and a generator with fixed voltage is located at the ``from" bus.
Fig.~\ref{fig: Nose_Curves} shows the power-voltage curves that result if the load's power factor is held fixed with several different amounts of reactive power injection.

As reactive support increases, the system can sustain larger increases in load before voltage collapse occurs. However, if reactive resources are used to maintain voltages at their nominal levels (1 p.u.), the load margin\textcolor{black}{, as measured from the operating voltage of 1 p.u. to the point of bifurcation, decreases. This is clearly seen in Fig. \ref{fig: Nose_Curves} by the convergence of the arrows associated with the $\boldsymbol *$ symbols (bifurcation) and the $\bullet$ symbols (nominal operating voltage) as reactive support increase.} On the other hand, as load increases, the magnitude of the derivative of the PV curve (with respect to load) increases, suggesting that this derivative is a useful indicator of proximity to the bifurcation.

If load varies stochastically with known variance $\sigma_{P_{D}}^{2}$, (\ref{eq:Var_Vt}) describes the expected voltage variance \textcolor{black}{via the delta method~\cite{Oehlert:1992}} at the load bus:
\begin{align}
\sigma_{\mathrm{V}_{t}}^{2}\approx\left(\left.\frac{d\mathrm{V}_{t}}{dP_{D}}\right|_{\mathrm{E}[P_{D}]}\right)^{2}\!\sigma_{P_{D}}^{2}\label{eq:Var_Vt},
\end{align}
where $\mathrm{E}[P_{D}]$ is the expected value of the load. Predicting the distance to static voltage collapse with variance measurements can be accomplished by (i) drawing the PV curve for a system, (ii) defining a loading margin \textcolor{black}{(in terms of complex power $P\!+\!jQ$)} on the curve which should not be exceeded, and then (iii) calculating the expected bus voltage variance at this threshold. If measured voltage variance exceeds this threshold value then the system may be at risk of exceeding its stability limits. If reactive support is high, the voltage magnitude of the system may be an unreliable real time measure of voltage stability. The bus voltage variance statistic, however, can potentially tell a more complete story about system stability. In the following sections, the stability information encoded in the variance is leveraged in order to make real-time, data-driven control decisions.

\subsection{System Model Overview}
A stochastically forced power system can be modeled with a set of Differential-Algebraic Equations (DAEs) of the form
\begin{align}
\dot{\mathbf{x}} &= \mathbf{f}(\mathbf{x},\mathbf{y})\label{eq:Diff_Eqs} \\
\mathbf{0} &= \mathbf{g}(\mathbf{x},\mathbf{y},\mathbf{u}(t))\label{eq:Alg_Eqs}
\end{align}
where $\mathbf{f}$, $\mathbf{g}$ represent the differential and algebraic systems, $\mathbf{x}$, $\mathbf{y}$ are the differential and algebraic state variables, and $\mathbf{u}(t)$ represents the time-varying stochastic (net) load fluctuations~\cite{Ghanavati:2015,Amini:2016,Milano:2010}.
Neglecting for the moment the slow changes in load level, the complex load at time $t$ can be represented by (\ref{eq: S(t)}):
\begin{equation}
\mathbf{S}(t)=\mathbf{S}_{0}(1+\mathbf{u}(t))\label{eq: S(t)}
\end{equation}
with the dynamics of the fast load fluctuations given by the Ornstein-Uhlenbeck process expressed in (\ref{eq: u_dot}):
\begin{equation}
\mathbf{\dot{u}}=-E\mathbf{u}+\textcolor{black}{\Sigma}\underline{\xi}\label{eq: u_dot}
\end{equation}
where $E$ is a diagonal matrix of inverse time correlations, $\underline{\xi}$ is a vector of zero-mean independent Gaussian random variables \textcolor{black}{whose standard deviations are given on the diagonal of the $n$x$n$ diagonal matrix $\Sigma$. This paper} assumes that a grid operator can estimate the statistics of load fluctuations ($E$ and \textcolor{black}{$\Sigma$}) from measurements.

\subsection{Computing the Algebraic Variable Covariance Matrix}
The process for deriving the approximate covariance matrix for all variables in a stochastically forced power system is derived in \cite{Ghanavati:2015}. This computation allows one to characterize the statistics of a system that is approaching a bifurcation. This method is based on linearizing the equations encompassed by (\ref{eq:Diff_Eqs}) and (\ref{eq:Alg_Eqs}) and then algebraically solving for $\Delta\mathbf{\dot{x}}$ and $\Delta\mathbf{\dot{u}}$ by eliminating the algebraic variable vector $\Delta\mathbf{y}$:
\begin{equation}
\left[\begin{array}{c}
\Delta\mathbf{\dot{x}}\\
\Delta\mathbf{\dot{u}}
\end{array}\right]=\left[\begin{array}{cc}
\mathrm{A}_{s} & -\mathbf{f_{y}g}_{\mathbf{y}}^{-1}\mathbf{g_{u}}\\
0 & -E
\end{array}\right]\left[\begin{array}{c}
\Delta\mathbf{x}\\
\Delta\mathbf{u}
\end{array}\right]+\left[\begin{array}{c}
0\\
\textcolor{black}{\Sigma}
\end{array}\right]\underline{\xi}.\label{eq: del_x_u_dot}
\end{equation}
where $\mathrm{A}_{s}$ is the standard state matrix. Using \noindent $\mathbf{z}=\left[\begin{array}{cc}
\Delta\mathbf{x} & \Delta\mathbf{u}\end{array}\right]^{\top}$,~\ref{eq: del_x_u_dot}) can be rewritten with compact matrices $A$ and $B$ via
\begin{equation}
\dot{\mathbf{z}}=A\mathbf{z}+B\underline{\xi}.\label{eq: z_dot}
\end{equation}
As introduced in \cite{Gardiner:2012}, the Lyapunov equation (\ref{eq: Lyap_Cov}) can be solved numerically\footnote{\textcolor{black}{Singularity of the state matrix is required for (\ref{eq: Lyap_Cov}) to have a solution. As the system approaches a singularity-induced bifurcation and ${\rm A}_s$ approaches singularity, the predicted variance will approach infinity.}} to calculate the covariance matrix of $z$, where $A$ and $B$ are defined in~(\ref{eq: z_dot}):
\begin{equation}
A\sigma_{\mathbf{z}}^{2}+\sigma_{\mathbf{z}}^{2}A^{\top}=-BB^{\top}.\label{eq: Lyap_Cov}
\end{equation}
Since \textcolor{black}{the linearized output} $\Delta\mathbf{y}$ is given by \noindent $\Delta\mathbf{y}=K\Delta\mathbf{z}$ where $K\equiv\left[\begin{array}{cc}
-\mathbf{g}_{\mathbf{y}}^{-1}\mathbf{g_{x}} & -\mathbf{g}_{\mathbf{y}}^{-1}\mathbf{g_{u}}\end{array}\right]$, the state variable covariance matrix can be transformed into the algebraic variable covariance matrix via
$\sigma_{\mathbf{y}}^{2}=K\sigma_{\mathbf{z}}^{2}K^{\top}\label{eq: Cov_y}$. A subset of the diagonal entries of $\sigma_\mathbf{y}^2$ contain the bus voltage variances.

\subsection{Adapting HELM to Solve CPF}
The Continuation Power Flow (CPF) problem is a classic approach to understanding and predicting voltage instability. As outlined in~\cite{Ajjarapu:1992}, CPF involves drawing PV curves given load and generation increase rates
using iterative Newton-Rapson methods.
As introduced in~\cite{Trias:2012}, iterative techniques, such as Newton-Raphson, can encounter a numerical issues, such as divergence or finding undesirable low-voltage solutions,
when solving the nonlinear power flow equations, particularly when a system approaches a Saddle-node bifurcation.
An alternative is to use the Holomorphic Embedding Load-flow Method (HELM), which uses complex analysis and recursive techniques to overcome these numerical difficulties. If one exists, HELM is guaranteed to compute the high voltage power flow solution~\cite{Rao:2015}.

Prior work~\cite{Subramanian:2015} provides an important foundation for using HELM to solve for the static stability margins of a power system. After generating the holomorphic voltage functions, the largest, positive zero of the numerator of the Pad\'{e} approximant approximates the maximum power transfer point of the system. This method, though, scales all loads at uniform rates, and does not account for more than one single generator bus in the system. In order to solve these problems, we derive a new method for scaling loads from a known base case solution. This approach allows loads and generators to scale at different rates.

In the conventional CPF problem, generation participation rates are assigned to generators to pick up excess load as it is scaled. This is not the approach we took. For mathematical simplicity, we instead solve the base case power flow solution and then fix the generator voltage phase angles. As load increases at the load buses, generation throughout a system increases quasi-proportionally to the electrical distance between the generator and the load. Electrically proximal generators respond with the largest generation increases, while electrically distant generators respond with  smaller increases; we justify these \textcolor{black}{simplifying} assumptions in~\cite{Chevalier:2016}. \textcolor{black}{Incoporating droop-coefficient-based generator loading rates remains for future work. 
}

\textcolor{black}{We originally derived the full details of the method in~\cite{Chevalier:2016}. The mathematics are too lengthy to be shown in this paper, but they are summarized in the remainder of this subsection. We begin by defining a holomorphic voltage function for the $i^{\rm th}$ power system bus voltage via the following power series:
\begin{align}
V_{i}(s) & =\sum_{n=0}^{\infty}V_{i}[n]s^{n},\label{eq: V_PS}
\end{align}
where the variable $s$ is a complex holomorphic function parameter. The $s=0$ condition yields the complex system wide voltages for a given base case power flow solution (which may be solved for via HELM or Newton-Raphson). If this power flow solution is known, $V_i[0]$ is known $\forall i$ in the system. With this definition,} the holomorphically embedded power flow equation at the $i^{\mathrm{th}}$ PQ bus in an $N$ bus power system may be stated:
\begin{align}
\sum_{k=1}^{N}Y_{i,k}V_{k}(s)=\frac{S_{i}^{*}+sk_{i}S_{i}^{\mathrm{*}}}{V_{i}^{*}(s^{*})},\quad i\in\mathrm{PQ}. \label{eq:HEML1}
\end{align}
\textcolor{black}{Equation (\ref{eq:HEML1}) has the following attributes:
\begin{itemize}
\item The holomorphic parameter $s$ scales the load as it is increased from $s=0$. If $s$ is real, the power factor of the load is constant as apparent power scales.
\item $S_i=P_i+jQ_i$ is the complex power injection at bus $i$.
\item The exponent $^*$ denotes complex conjugation.
\item The parameter $k_{i}$, which can be positive, negative, or 0, corresponds to the rate at which bus $i$ will be loaded as $s$ increases from 0. If $k_{i}=0$, the load at bus $i$ will not change as $s$ increases.
\end{itemize}}
The holomorphically embedded equations at voltage controlled buses (PV and reference) are given by
\begin{equation}
V_{i}(s)=\mathrm{V}_{i}e^{j\theta_{i}},\quad i\in\left\{ \mathrm{PV}\cup\mathrm{r}\right\}. \label{eq: PV_SW_V}
\end{equation}
Generator voltages are independent of $s$ since reactive power limits are not considered in this formulation. \textcolor{black}{With the structure given by (\ref{eq: V_PS}), (\ref{eq:HEML1}), and (\ref{eq: PV_SW_V}), the formulations given in~\cite[eqs. (4.65)-(4.89)]{Chevalier:2016} may be used to recursively compute the power series coefficients of $V_i(s)$. Once done, the complex bus voltage at PQ bus $i$ for some arbitrary loading level $s=s_l$ may be computed via
\begin{align}
{\rm V}_{i}e^{j\theta_{i}}=\left.\sum_{n=0}^{\infty}V_{i}[n]s^{n}\right|_{s=s_{l}}.
\end{align}
If the voltage functions are evaluated from $s=0$ to the voltage collapse load $s=s_c$, the CPF voltage magnitude curves may be drawn analytically. In order to identify the critical load $s=s_c$, the formulations in~\cite[eqs. (3.46)-(3.50)]{Chevalier:2016} may be used to generate the Pad\'{e} approximants $A$ and $B$ of the power series $V_i(s)$:
\begin{equation}
\sum_{n=0}^{\mathrm{N_{c}}-1}V[n]\left(s^{n}\right) = \frac{\sum_{n=0}^{\frac{\mathrm{N_{c}}-1}{2}}A[n]\left(s^{n}\right)}{\sum_{n=0}^{\frac{\mathrm{N_{c}}-1}{2}}B[n]\left(s^{n}\right)}.\label{eq: Holo=Pade}
\end{equation}
From this formulation, $s_c$ will be equal to the smallest, positive real root of $A$~\cite{Subramanian:2015}.}

\begin{figure}
\noindent \begin{centering}
\includegraphics[scale=0.44]{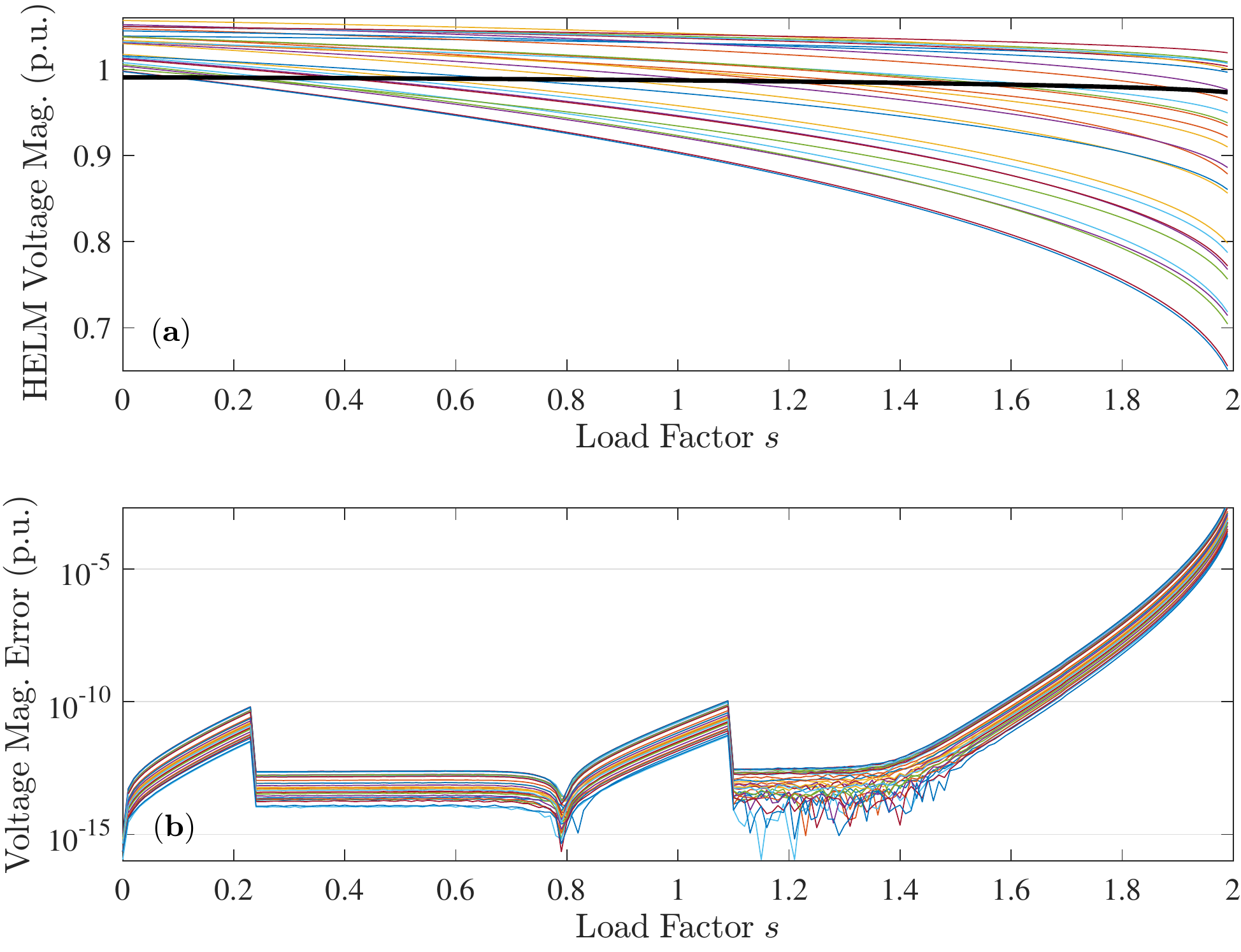}
\par\end{centering}
\caption{\label{fig: HELM_Voltage}Panel (${\rm {\bf a}}$) shows the voltage magnitude for load buses of the 39 bus system as $s$ is scaled, as computed by our adaptation of HELM; the curves in this panel are drawn analytically. The thick black curve is the voltage magnitude of bus 20, whose load is decreasing as $s$ increases, according to equation (\ref{eq: k_i}). Panel (${\rm {\bf b}}$) shows the voltage magnitude difference (error) for each PV curve between the NRPF and HELM solutions. The error is numerically insignificant until $s$ approaches the bifurcation point ($s\approx 2$).}
\end{figure}

In order to validate this continuation method, which we refer to as \emph{CPF via HELM}, we test our method on the IEEE 39 bus system. We first define the vector $\mathbf{k}$ which has length 39. The elements of this vector contain the respective arbitrary loading rates of the buses in the system.
\begin{equation}\label{eq: k_i}
\mathbf{k}_{i}=\left\{ \begin{array}{ccc}
1 &  & i\in\{3,\:4,\:7,\:8\}\\
-0.2 &  & i=20\\
0 &  & \textrm{otherwise}
\end{array}\right.
\end{equation}
The loads of the system scale in the following way, where $S_{0_i}$ is the base load at bus $i$ in the system:
\begin{align}
S_{i}=S_{0_i}+s\mathbf{k}_{i}S_{0_i}.\label{eq: s_scales}
\end{align}
In this example, we scale $s$ from 0 to 1.99, at which point the Saddle-Node Bifurcation occurs. Once the  Pad\'{e} approximants are known for each bus, we scale $s$ and solve for the complex voltages at each bus. The resulting PV curves are shown in Panel (${\rm {\bf a}}$) of Figure~\ref{fig: HELM_Voltage}.

We also validated HELM against conventional Newton Raphson Power Flow (NRPF). As $s$ was increased and the load was scaled via HELM, we solved for load bus voltages using NRFP. We then plotted the difference between the HELM and the NRPF voltage magnitudes in panel (${\rm {\bf b}}$). These results suggest that \emph{CPF via HELM} computes load bus \textcolor{black}{complex voltages with a very low degree of error} for a given level of load increase. \textcolor{black}{Accordingly, the critical loading levels computed by the numerator of (\ref{eq: Holo=Pade}) are a sufficiently accurate 
approximation for the exact voltage collapse load values.}

\subsection{Deriving a Probabilistic Loading Margin from First Passage Processes}

\begin{figure}
\noindent \begin{centering}
\includegraphics[viewport=120 210 600 320,clip,scale=0.48]{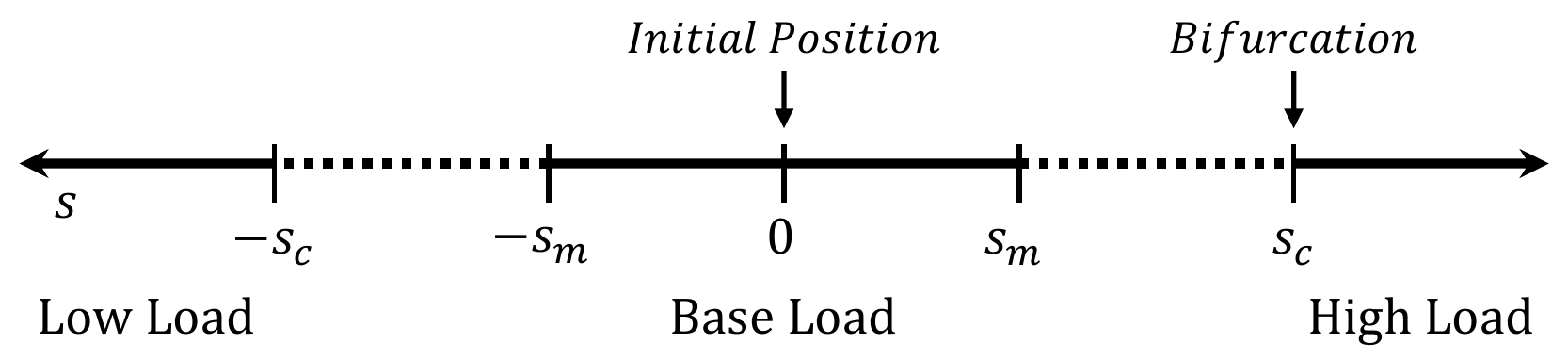}
\par\end{centering}
\protect\caption[First Passage Diagram]{\label{fig: First Passage Diagram} An illustration of the values that the holomorphic parameter $s$, which starts at a base load $s=0$, can attain during its random walk. $s_m$ is the load level at which the probability of $s$ hitting the Saddle-Node bifurcation $s_c$ exceeds a probability limit.}
\end{figure}

CPF (via HELM or NRPF) allows one to estimate how much \textcolor{black}{of a load} margin exists between a current operating point and voltage instability. However, it does not \textcolor{black}{compute the \textit{probability} that a particular system will destabilize due to stochastic load buildup}. This section builds on the First Passage Process literature to systematically compute the probability that load will not increase beyond a collapse threshold during a given time period. To do so, \textcolor{black}{we consider the holomorphic parameter $s$ from (\ref{eq:HEML1}) which will scale the base complex power of the $i^{\rm th}$ bus according to (\ref{eq: s_scales}). To capture slow stochastic load changes, the evolution of} parameter $s$ is modeled as a Wiener Process in which $s$ begins at the origin and takes Gaussian-distributed steps with variance $2D$:
\begin{align}
s[0] &= 0\label{eq: RW1} \\
s[k+1] &= s[k] + \sqrt{2D}\cdot\mathcal{N}(1,0).\label{eq: RW2}
\end{align}
The values which $s$ may attain are shown in Figure~\ref{fig: First Passage Diagram}, where $s_c$ corresponds to a Saddle-Node bifurcation of the algebraic power flow equations. \textcolor{black}{More explicitly, if the complex load at some bus reaches $S=S_0(1+s_c)$, then the system's voltage will collapse (for notational simplicity, non-unity loading rates are not taken into account).} If $s$ is allowed to drift over a time period $\textcolor{black}{\Delta} t$, and an absorbing boundary condition (the point of collapse) sits at $s=s_c$, the survival probability \textcolor{black}{(SP)} of the system may be computed. \textcolor{black}{The SP refers to the probability that that the system will \textit{not} experience voltage collapse due to the wandering of $s$ during $\Delta t$}. As derived in \cite{Redner:2001}, the \textcolor{black}{SP} for a system starting at $s=0$ may be estimated as
\begin{align}
\textcolor{black}{\rm{SP}} & \approx \mathrm{erf}\left(\frac{s_c}{\sqrt{4D\textcolor{black}{\Delta}t}}\right)\label{eq: Survival_Probability},
\end{align}
where $\mathrm{erf}$ is the error function, and $D$ is the diffusion coefficient \textcolor{black}{of (\ref{eq: RW2})} which is based on load variability. Equation~(\ref{eq: Survival_Probability}) gives the probability that the parameter $s$ will not cross the voltage collapse threshold $s_c$ at any point during time $\textcolor{black}{\Delta}t$.

\begin{figure}[H]
\noindent \begin{centering}
\includegraphics[scale=.45]{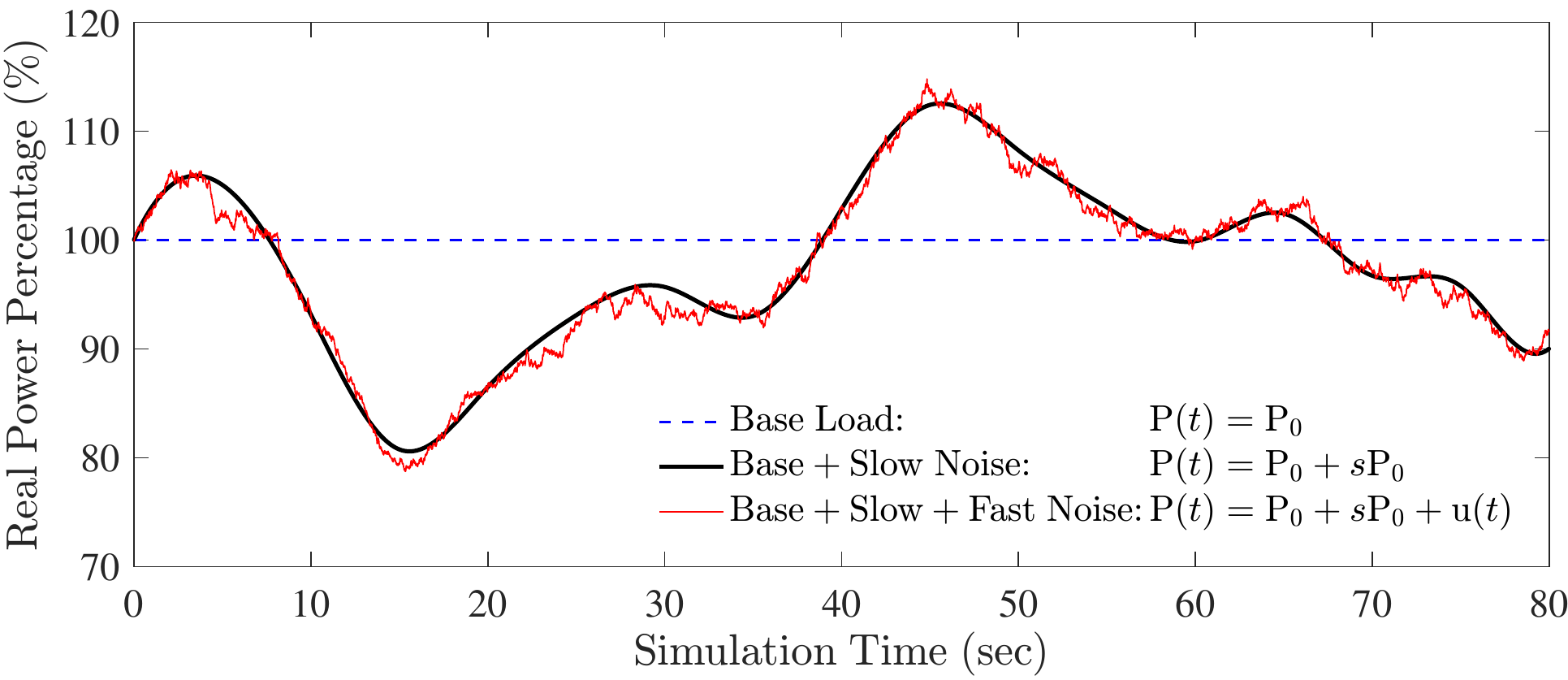}
\par\end{centering}
\protect\caption{\label{fig: Fast_Slow_Noise} The interaction between base load, the fast noise ${\rm u}(t)$  of the Ornstein-Uhlenbeck process, and the slow noise from the randomly walking Holomorphic scaling parameter $s$ is portrayed. }
\end{figure}

Additionally, we introduce the value $s_m$, where \textcolor{black}{$0<s_m<s_c$}. This scalar value corresponds to the maximum allowable load level \textcolor{black}{$S=S_0(1+s_m)$} that can be reached before some \textcolor{black}{operator specified} probability of voltage collapse grows too high. \textcolor{black}{Said differently, $s_m$ is chosen such that if the load starts diffusing from $S=S_0(1+s_m)$, the probability that it will \textit{not} reach a voltage collapse load of $S=S_0(1+s_c)$ will be equal to operator specified survival probability $\rm SP^*$. For example, if an operator specifies that the system must have a survival probability of at least $\rm SP^*$ over future time interval $\Delta t$, then $S_0(s_c-s_m)$ will be the load margin between the specified probabilistic threshold and true voltage collapse:
\begin{align}
{\rm SP^*} & \approx \mathrm{erf}\left(\frac{s_c-s_m}{\sqrt{4D{\Delta}t}}\right),\label{eq: SP_approx}
\end{align}
where $D$ is a diffusion coefficient that sets the standard deviation of the stochastic load deviations per unit time. 
As a result, the probability of voltage collapse ${\rm VCP^*}$ is given by:
\begin{align}
{\rm VCP^*} & =1-{\rm SP^*} \label{eq: VCP_approx}.
\end{align}}
The connection between first passage processes and voltage collapse is further detailed in \cite{Chevalier:2016}. We assume that system load changes will incorporate both fast and slow load changes on top of some base load condition. The interaction between these fluctuations and the base load are illustrated in Fig.~\ref{fig: Fast_Slow_Noise}.

\section{Controller Design}\label{Controller Outline}

This section introduces a Variance Based Controller (VBC), which is subsequently shown to successfully mitigate the probability of voltage collapse. For benchmarking and illustration purposes, we also introduce two other, more conventional, controllers: a Mean Based Controller (MBC) and a Reference Based Controller (RBC). For clarity, we introduce these controllers in reverse order of complexity (least to most). In actual implementation, both the MBC and VBC controller systems require real time (PMU) load voltage observability and a controllable reactive power resource, such as a Synchronous Condenser or a Static VAR Compensator (SVC) that can support load voltage.

\subsection{Reference Based Controller Overview}

\begin{figure}[H]
\noindent \begin{centering}
\includegraphics[scale=1.4]{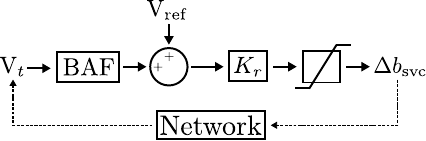}
\par\end{centering}
\protect\caption{\label{fig: RBC_Model} Reference Based Controller (RBC). Local terminal bus voltage ${\rm V}_t$ is the only feedback signal.}
\end{figure}

The RBC does not rely on the Wide Area Measurement System (WAMS); instead, it uses a local voltage terminal measurement ${\rm V}_t$ as a feedback signal to control the reactive power injected by a ``quasi-static SVC" device\footnote{In typical power system modeling, SVC devices can be dynamically modeled with sets of ODEs. Since we are using the statistics of buffered time series data to make control decisions, we have the SVC take discrete, rather than continuous, control action every $T_w$ (time window) seconds. We therefore refer to the device as a ``quasi-static SVC".} This relatively simple approach to feedback SVC control is illustrated in Fig.~\ref{fig: RBC_Model}. In this diagram, $\Delta b_{\rm svc}$ is the change in susceptance at the SVC and the regulator gain $K_r$ is tuned to properly correlate voltage changes with reactive power injections. The reactive power changes are limited by the size, in MVAr, of the SVC. The ``BAF'' block represents a Buffered Average Filter that provides a rolling average bus voltage magnitude over an operator-specified time window $T_w$. After $T_w$ seconds, the BAF computes average voltages and the SVC adjusts its reactive power injection as needed. Finally, the ``Network" block represents the physical feedback provided by the natural evolution of bus voltages due to control input, load fluctuations, and system dynamics.

\subsection{Mean Based Controller Overview}
\begin{figure}[H]
\noindent \begin{centering}
\includegraphics[scale=1.3]{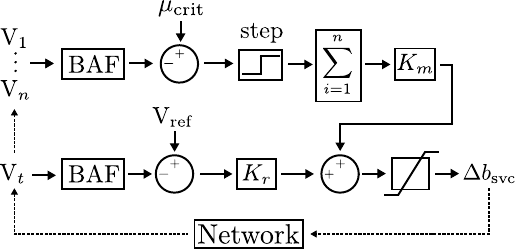}
\par\end{centering}
\protect\caption{\label{fig: MBC_Model} Mean Based Controller (MBC). The WAMS acquire voltage magnitude data ${\rm V}_1 \cdots {\rm V}_n$ to use as feedback signals for the SVC.}
\end{figure}

Similar to the Automatic Voltage Control (AVC) system outlined in~\cite{Liu:2010}, the MBC relies not only on local terminal voltage, but also on bus voltage magnitude data from a WAMS. These data, sampled at 30Hz, are also passed through a BAF with time window $T_w$ and then are each subtracted from some critical magnitude $\mu_{\rm crit}$ and summed together. The thresholds imposed by $\mu_{\rm crit}$ are chosen based on minimum tolerable voltages (such as 0.98 p.u., for example); probabilistic security margins are not considered. As illustrated in Fig.~\ref{fig: MBC_Model}, the voltage magnitudes ${\rm V}_1 \cdots {\rm V}_n$ represent WAMS data from PMUs, and the gain $K_m$ is set based on how the operator wishes for the WAMS feedback and the local feedback to interact.
The ``step'' function in Fig.~\ref{fig: MBC_Model} (and Fig.~\ref{fig: VBC_Model}) operates exactly as the unit step function of equation (\ref{eq: US}). Its purpose is to ensure that only WAMS bus voltages that are lower than the critical voltage $\mu_{\rm crit}$ impact the feedback signal.
\begin{equation}\label{eq: US}
x\cdot u(x)=\left\{ \begin{array}{c}
x\quad x\geq0\\
0\quad x<0
\end{array}\right.
\end{equation}
\textcolor{black}{In the context of Fig. \ref{fig: MBC_Model}, $x_i=\mu_{\rm crit}-{\rm BAF}({\rm V}_i)$}.

\subsection{Variance Based Controller Overview}
\begin{figure}
\noindent \begin{centering}
\includegraphics[scale=1.3]{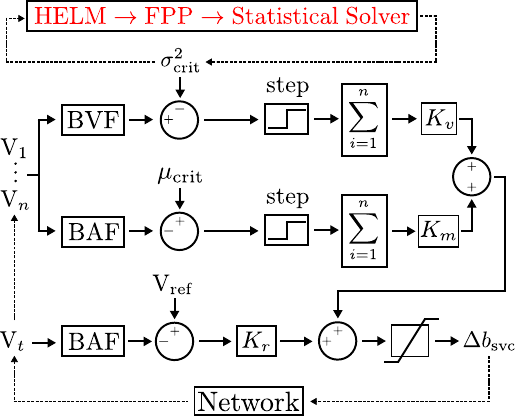}
\par\end{centering}
\protect\caption{\label{fig: VBC_Model} Variance Based Controller (VBC). Wide Area Measurement Systems (WAMS) gather load pocket voltage magnitude data. Buffered Average Filters (BAFs) and Buffered Variance Filters (BVFs) are used to quantify bus voltage magnitude and variance. The step functions ensure that \textbf{only} critically low magnitude and critically large variance measurements have effect on $b_{\rm SVC}$.}
\end{figure}

The VBC builds on the tools described in Section~\ref{Background} in the following way. CPF via HELM is used to quickly determine how much the load within a load pocket of concern may increase before the system undergoes static voltage collapse. Next, the First Passage Process is used to determine the load level below which the probability of voltage collapse is sufficiently low.
Using this loading level, the critical bus voltage variances are found by leveraging the analytical covariance matrix solver along with load noise estimation. Finally, these critical variances are used as a feedback signal to control the reactive power injected by the quasi-static SVC device, as shown in Fig.~\ref{fig: VBC_Model}.

The VBC process is formally described in Algorithm~\ref{alg: VBC}. The BVF, or Buffered Variance Filter, is similar to the BAF in that it computes the variance from a window of measurement data. The constant $K_v$ is a feedback gain parameter for the variance measurements, and is tuned to allow the controller to use both the voltage magnitude and the variance feedback signals.

\begin{algorithm}\label{alg: VBC}
\caption{Variance Based Controller (VBC)}
\textbf{START}\\
\begin{enumerate}[label=\textbf{\arabic*},start=1]
	\item Perform CPF via HELM on load pocket\\
	\item Determine voltage collapse loading factor $s_c$\\
	\item Based on desired probabilistic security margin,\\
          determine loading factor $s_m$ s.t. $0<s_m<s_c$\\
	\item Computationally scale loads based on $s_m$ and then\\
    analytically solve for load pocket critical variance
	\item Use critical variances and magnitude constraints as inputs to Fig.~\ref{fig: VBC_Model} controller
    \end{enumerate}
 \eIf{New State Estimate Data Available}{
 Return to \textbf{START}}{Return to $\textbf 5$}
\end{algorithm}


\section{3 Bus System Illustration}\label{Test Results}
In order to illustrate and compare the effectiveness of the controllers, we test each (RBC, MBC and VBC) on a three bus system with \textcolor{black}{identical simulation parameters between tests. In order to provide a complete system description, the simulation, control, and data files have been publicly posted online\footnote{\textcolor{black}{https://github.com/SamChevalier/VoltageCollapse3Bus}} for open source access}.

\subsection{System Overview}
\begin{figure}
\noindent \begin{centering}
\includegraphics[scale=.9]{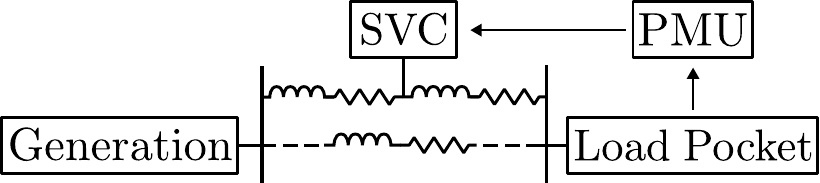}
\par\end{centering}
\protect\caption{\label{fig: 3_Bus_System} 3 Bus test case. Aggregate generation (bus 1)  feeds an aggregate load pocket (bus 2) with voltage supported by a local SVC (bus 3).}
\end{figure}

In our three-bus test case (Fig.~\ref{fig: 3_Bus_System}), aggregate generation is connected to a heavily loaded aggregate load pocket (such as a city). The voltage magnitude of this load pocket is supported by a fully controllable quasi-static SVC, and a PMU feeds voltage magnitude data back to the SVC in real time. In the case of the RBC, these PMU data are neglected.

We outfit the generator with a $4^{\rm th}$ order Synchronous Machine (SM), a $4^{\rm th}$ order Automatic Voltage Regulator (AVR), and a $3^{\rm rd}$ order Turbine Governor (TG). System parameters are approximately based on the WSCC 9-bus system, and component models are those described in~\cite{Milano:2010}. At the SVC, we choose a buffering time window of $T_w=3{\rm s}$. The load of the load pocket is constant power (PQ); the fast load fluctuations are described by the Ornstein-Uhlenbeck process of (\ref{eq: u_dot}) and the slow load variations are monotonically increased (see panel $(\bf b)$ of Fig.~\ref{fig: Test_Results}). Both fast and slow load fluctuations were applied to the active and reactive power demands equally in order to hold power factor constant.

To compare the three controllers, we (i) initialized the heavily loaded 3 bus system, (ii) performed a time domain simulation with a stochastically increasing load, and (iii) measured the survival time achieved by each controller. Fast Ornstein-Uhlenbeck noise is applied at each integration time step of $\Delta t = 0.01$. For each simulation, we record the random fast and slow noise vectors applied to the loads such that each controller experiences identical simulation realizations.

\begin{figure}
\noindent \begin{centering}
\includegraphics[scale=.45]{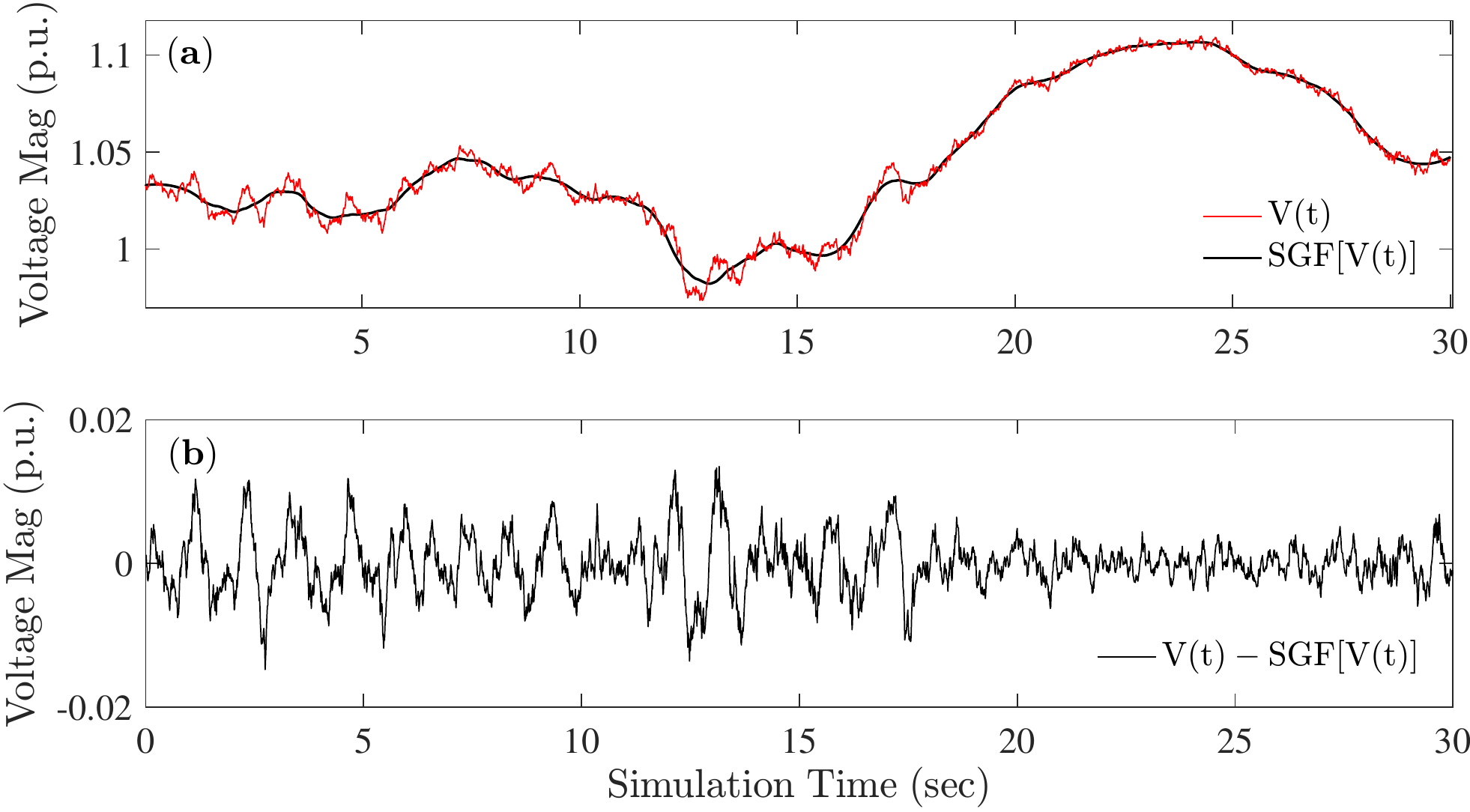}
\par\end{centering}
\protect\caption{\label{fig: Voltage_Detrend} Panel $(\bf a)$ shows a noisy time domain signal ${\rm V}(t)$ with slowly varying equilibrium changes (sped up for illustration purposes). A Savitzky-Golay Filter is applied to ${\rm V}(t)$. In panel $(\bf b)$, the filtered signal is subtracted from the noisy signal in order to generate the high frequency voltage fluctuations. This difference signal is used to compute bus voltage variance.}
\end{figure}

In order to estimate the high frequency variance $\sigma^2$ of a voltage signal whose underlying equilibrium point is constantly shifting due to the slow load fluctuations the real-time measurements must first be detrended. To do so, we employ a $2^{\rm nd}$ order FIR Savitzky-Golay Filter (SGF) to the voltage time series data and then subtract the smoothed voltage signal from the original data. This yields the zero-mean high frequency voltage perturbations, as illustrated in Fig.~\ref{fig: Voltage_Detrend}. 

\subsection{Simulation Results}
\begin{figure}
\noindent \begin{centering}
\includegraphics[scale=.45]{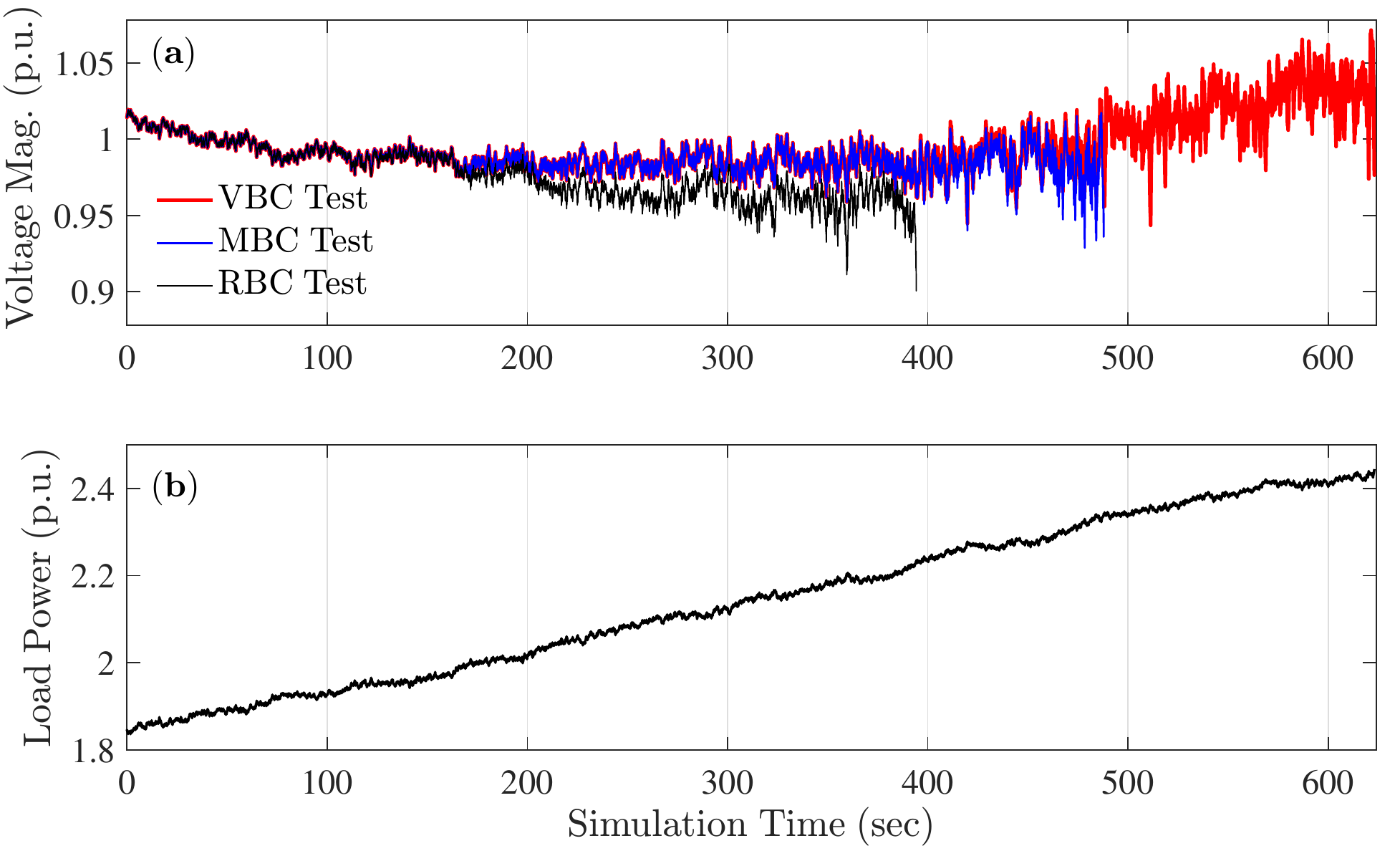}
\par\end{centering}
\protect\caption{\label{fig: Test_Results} Panel $(\bf a)$ shows the load bus voltage magnitude over the span of the simulations associated with all three controllers up to the point of voltage collapse. Panel $(\bf b)$ shows the active power demand at the load bus (identical for all three simulations).}
\end{figure}

With each controller, we simulated the system up until the point of voltage collapse. As previously indicated, the fast and slow noise vectors for the simulation were computed and saved before running each simulation, such that each controller experienced an identical simulation case. Table~\ref{tab: results} summarizes two primary test results: the amount of time each controller kept the system ``alive'' (prevented bifurcation) and the amount of load increase that the system was able to sustain.  Clearly, the Variance Based Controller most effectively preserved voltage stability while load increased.
\begin{table}[H]
\begin{centering}
\caption{Simulation Results Summary}\label{tab: results}
\begin{tabular}{cccc}
\textbf{Test Result} & \textbf{RBC} & \textbf{MBC} & \textbf{VBC}\tabularnewline
\hline 
Bifurcation Time (sec) & 394.26 & 487.87 & 623.05\tabularnewline
\hline 
Bifurcation Load Increase (\%) & 21.2\% & 27.0\% & 32.8\%\tabularnewline
\hline 
\end{tabular}
\par\end{centering}
\end{table}
To further illustrate these results, Fig.~\ref{fig: Test_Results} shows the load bus voltage magnitude over time (panel $(\bf a)$) for all three controllers until the point of bifurcation and the active power demand (panel $(\bf b)$) at the load bus.

The results in Fig.~\ref{fig: Test_Results} show that all three controllers take identical action until roughly 200 seconds. At this point, the \textcolor{black}{PMU feedback signal of the voltage magnitude from the load bus (bus 2 in Fig.~\ref{fig: 3_Bus_System})} begins to drop low enough to warrant control action. The RBC simulation bifurcates at around 400 seconds, but the MBC is able to maintain stability until about 490 seconds. At this point, the VBC begins to take control action due to the extreme increases in the bus voltage variance. Since it relies only on bus voltage magnitude data, the MBC is unaware that additional control action is needed and fails to maintain stability. 
\begin{figure}
\noindent \begin{centering}
\includegraphics[scale=.45]{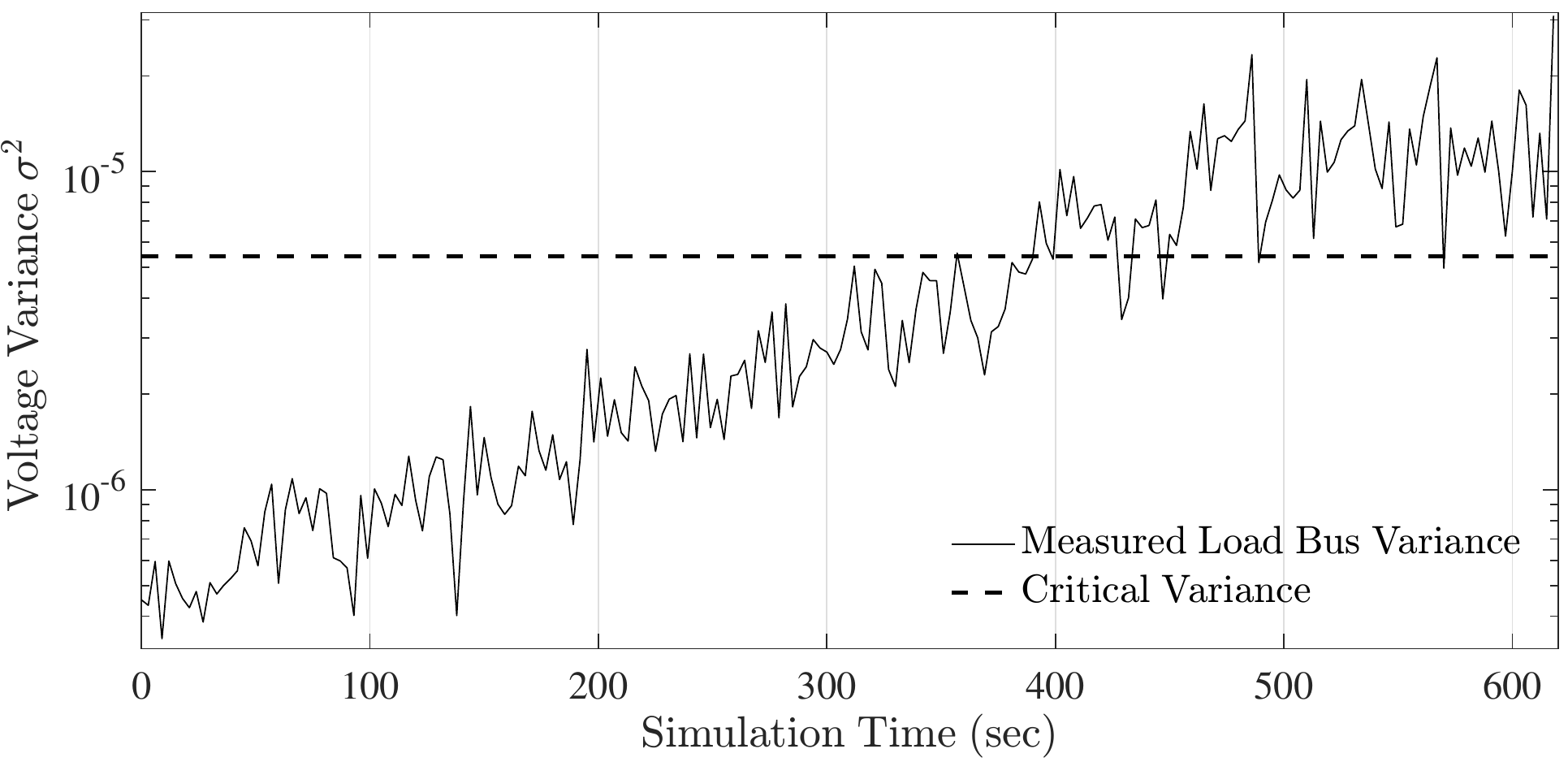}
\par\end{centering}
\protect\caption{\label{fig:Variance_Evolution} The discretely measured (every ${\rm T}_w=3$ seconds) load bus voltage variance is plotted over the simulation lifespan for the VBC.}
\end{figure}

In Fig.~\ref{fig:Variance_Evolution}, the bus voltage variance crosses the ``critical'' threshold just before $t=400$. The VBC simulation begins to call for increasing SVC support and thus prevents the system from bifurcating at $t=490$, when the MBC system fails. As can be inferred from Fig.~\ref{fig: Nose_Curves} and equation (\ref{eq:Var_Vt}), the bus voltage variance begins to show an exponential increase when the system load approaches the stability limit. As a result, the control signal associated with the bus voltage variance, $K_v(\sigma^2_{\rm meas} - \sigma^2_{\rm_crit})$, also begins to increase exponentially. This explains the upward trend of the bus voltage magnitude for the VBC test during the last 100 seconds of simulation (seen in panel $({\bf a})$ of Fig.~\ref{fig: Test_Results}).

\begin{figure}
\noindent \begin{centering}
\includegraphics[scale=.44]{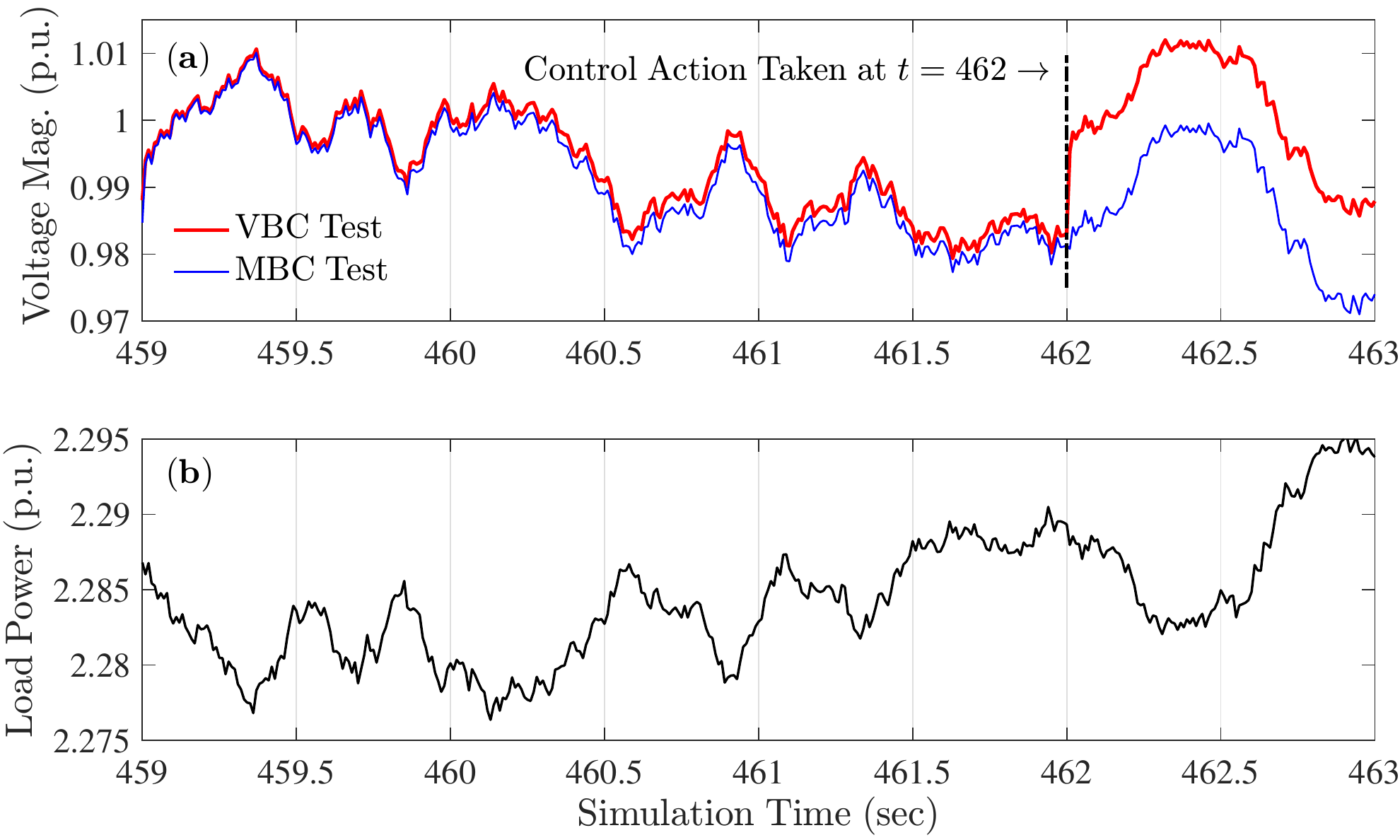}
\par\end{centering}
\protect\caption{\label{fig: Test_Results_Zoom} Panel $(\bf a)$ show the load bus voltage over a period of four seconds for two controllers, where both controllers taken control action at $t=462$ based on measurements taken over the time window of $t=459$ to $t=462$. Panel $(\bf b)$ shows the associated active power demand at the load bus.}
\end{figure}
It is helpful to consider a critical point when the VBC and the MBC take very different control actions. To do so, Fig.~\ref{fig: Test_Results_Zoom} zooms in on Fig.~\ref{fig: Test_Results} to the window of time from $t=459$ to $t=463$. In panel $(\bf b)$ of Fig.~\ref{fig: Test_Results_Zoom}, the load fluctuations from $t=459$ to $t=462$ spike downwards despite a slow upward trend. Since the system is operating close to the stability limit at this point, bus voltages spike high, above 1 per unit. Therefore, since the mean voltage over the time window from $t=459$ to $t=462$ appears relatively high, the MBC takes almost no control action. The VBC, on the other hand, measures an extremely high bus voltage variance and thus takes strong control action, despite the relatively high mean voltage magnitude (which is above 0.99 p.u.). This is but one of many examples of the VBC taking control action when the MBC does not. As more and more SVC support is added to the system, the mean voltage magnitude becomes an unreliable signal for system voltage health as the bifurcation voltage drifts closer to nominal system voltage. Bus voltage variance, on the other hand, is a robust indicator of a system's proximity to voltage collapse.



\section{39 Bus System Test Results}\label{Test Results 39}
\begin{figure}
\noindent \begin{centering}
\includegraphics[bb=60 90 700 520,clip,scale=0.35]{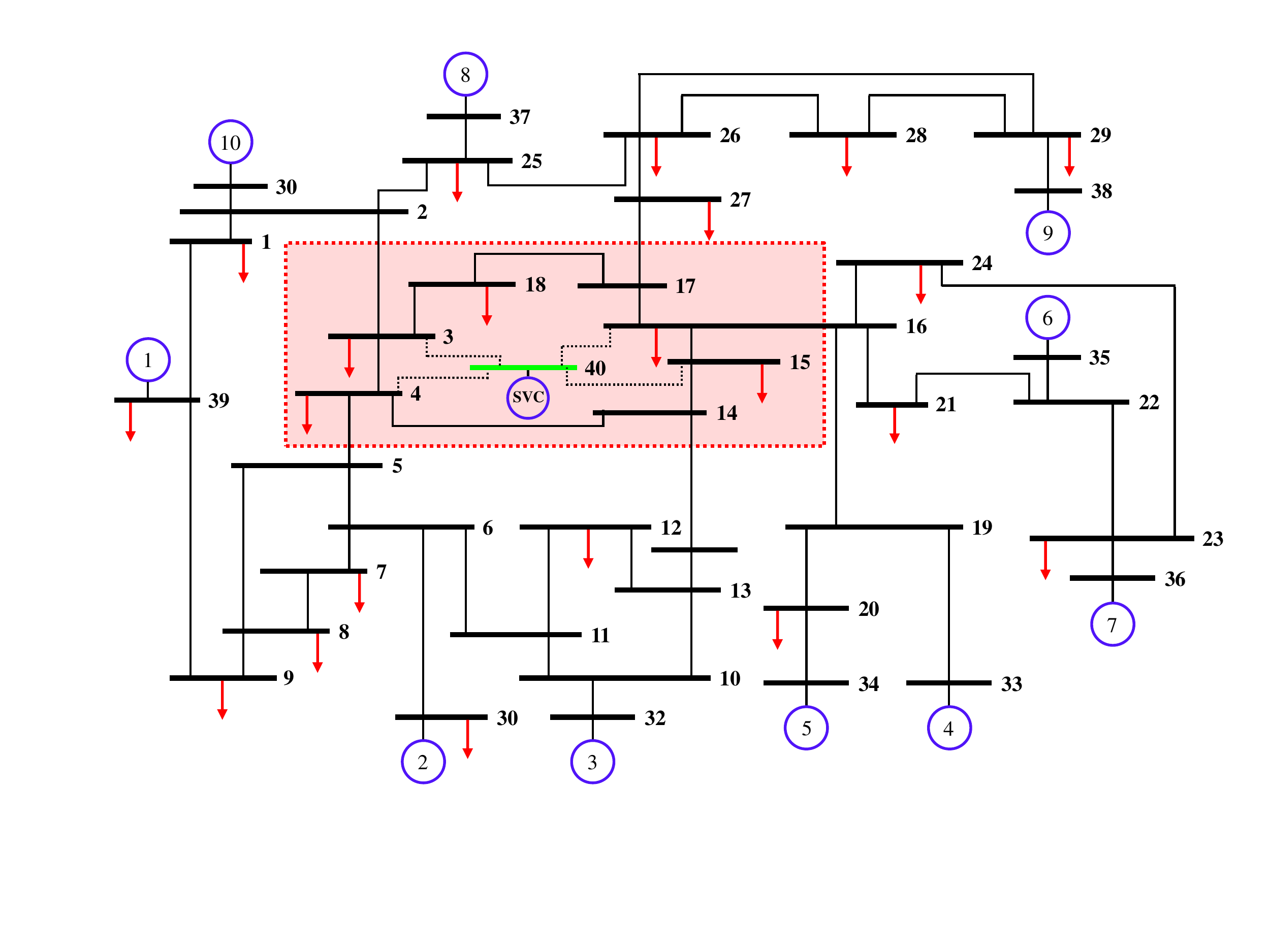}
\par\end{centering}
\protect\caption{\label{fig: 39_Bus_Pocket} IEEE 39 bus system with added SVC Bus.}
\end{figure}

For further validation, we tested the controllers on a modified version of the IEEE 39 bus system. As shown in Fig.~\ref{fig: 39_Bus_Pocket}, an SVC bus (bus ``40") was added to the system and connected to 4 other buses to form an observable (via PMU) load pocket with reactive support. To test the controllers in this system, monotonically increasing slow load changes were applied to all load pocket buses (3, 4, 14, 15, 16, 17, and 18), in addition to fast mean reverting Ornstein-Uhlenbeck load noise. As with the three-bus results in Fig.~\ref{fig: Test_Results}, the results clearly illustrate that the Variance Based Controller improves voltage stability most effectively, relative to the reference controllers.

\begin{figure}
\noindent \begin{centering}
\includegraphics[scale=0.45]{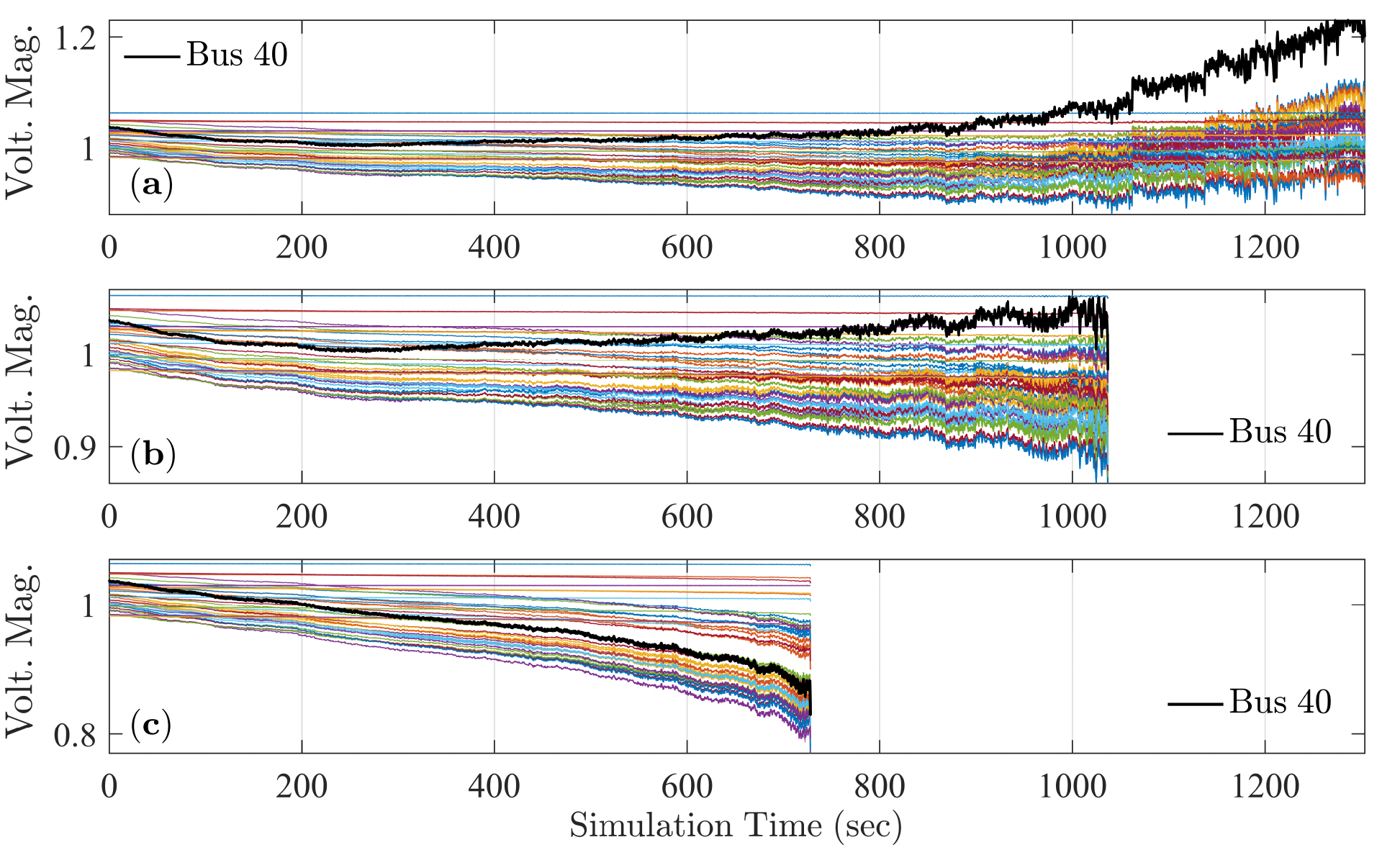}
\par\end{centering}
\protect\caption{\label{fig: Test_Results_39} Bus voltage magnitudes from simulations of the 
39 bus test case with three different control systems, as load increases up until the point of voltage collapse.
Panel (\textbf{a}) shows results from the VBC, panel (\textbf{b}) shows results from MBC, and panel (\textbf{c}) includes results for RBC. 
In each panel, the SVC bus voltage (bus 40) is noted.}
\end{figure}

\begin{figure}
\noindent \begin{centering}
\includegraphics[scale=0.45]{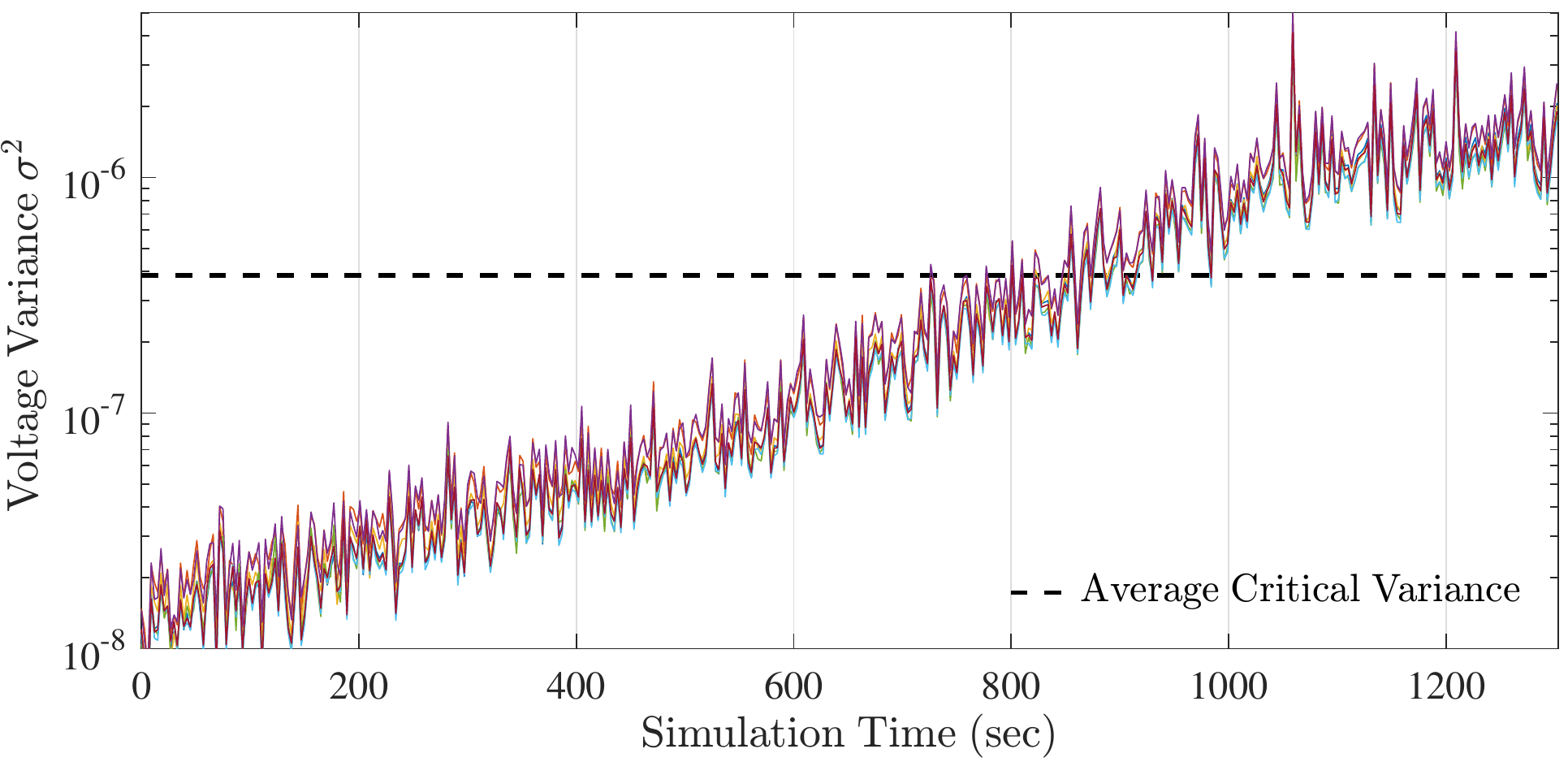}
\par\end{centering}
\protect\caption{\label{fig: Variance_Evolution_39} Bus voltage variance in the 39 bus test case, as load increases over time.}
\end{figure}

Fig.~\ref{fig: Test_Results_39} shows the voltage evolution for the tests corresponding with all three controllers. The VBC deters voltage collapse 270 seconds longer than the MBC and 579 second longer than the RBC. To better understand the success of the VBC, Fig.~\ref{fig: Variance_Evolution_39} shows the bus voltage variance and the average critical voltage variance. Since each bus has a unique critical voltage variance, as computed by ($\ref{eq: Cov_y}$), for the sake of graphical clarity, only the average critical variance is shown.

Because the VBC measures the differences between the measured and critical variances, and then scales these values by $K_v$ and sums them across buses, large increases in variance (which are expected as a system approaches its stability limit) lead to very large reactive power injections, even when voltage magnitude remains relatively ``high''. These variance increases are clearly seen in Fig.~\ref{fig: Variance_Evolution_39}.

\section{Conclusion}\label{Conclusion}
In this paper, we introduce and provide test results for a new reactive power control system that uses bus voltage variance as a control signal to improve voltage stability. Tests of this system on a three-bus test case show that the Variance Based Controller (VBC) can maintain voltage stability if load increases to 32.8\% above nominal, whereas a Mean Based Controller (MBC) allows for a load increase of only 27.0\% above nominal, and the Reference Based Control (RBC) allows for a load increase of only 21.2\%. Tests of the new control system on the 39 bus test case, in which load was steadily increasing, show that the VBC deterred voltage collapse 270 and 579 seconds longer than the MBC and RBC, respectively. Both sets of results clearly show that statistical information can be valuable in reducing the risk of voltage collapse.

Future work aims to extend the validation of the VBC to understand how it functions in the context of a larger system with more realistic load profiles. Similarly, the variance-based controller could be extended to include other types of statistical warning signs, such as autocorrelation. \textcolor{black}{In order to provide formal performance guarantees for the proposed statistical control system, there is a need for additional studies to describe the conditions under which including voltage variance (and other statistical) feedback in a reactive power control system leads to improved voltage control performance and stability. Additionally, it would be useful to reformulate CPF via HELM to incorporate generation increase rates, derived from droop control settings, in the holomorphic voltage functions for PV buses from (\ref{eq: PV_SW_V}). The incorporation of these generation increase rates could allow for the computation of an even more realistic load margin and thus better variance predictions. Finally, in moving towards a more practical implementation of these methods, future work aims to understand the interaction between the controllers developed in this paper and the other mechanisms which contribute to voltage collapse such as overexcitation limiters (OELs) and on-load tap changing (OLTC) transformers.}

\bibliographystyle{ieeetr}
\bibliography{Var_and_Control}

\begin{IEEEbiography}[{\includegraphics[width=1in,height=1.25in,clip,keepaspectratio]{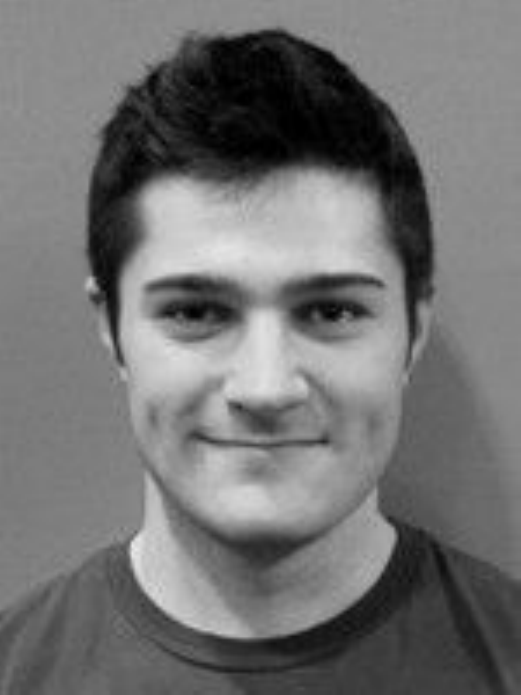}}]{Samuel C.~Chevalier} (S`13) received M.S. (2016) and B.S. (2015) degrees in Electrical Engineering from the University of Vermont, and he is currently pursuing the Ph.D. in Mechanical Engineering from MIT. His research interests include stochastic power system stability, renewable energy penetration, and Smart Grid applications.
\end{IEEEbiography}

\begin{IEEEbiography}[{\includegraphics[height=1.25in]{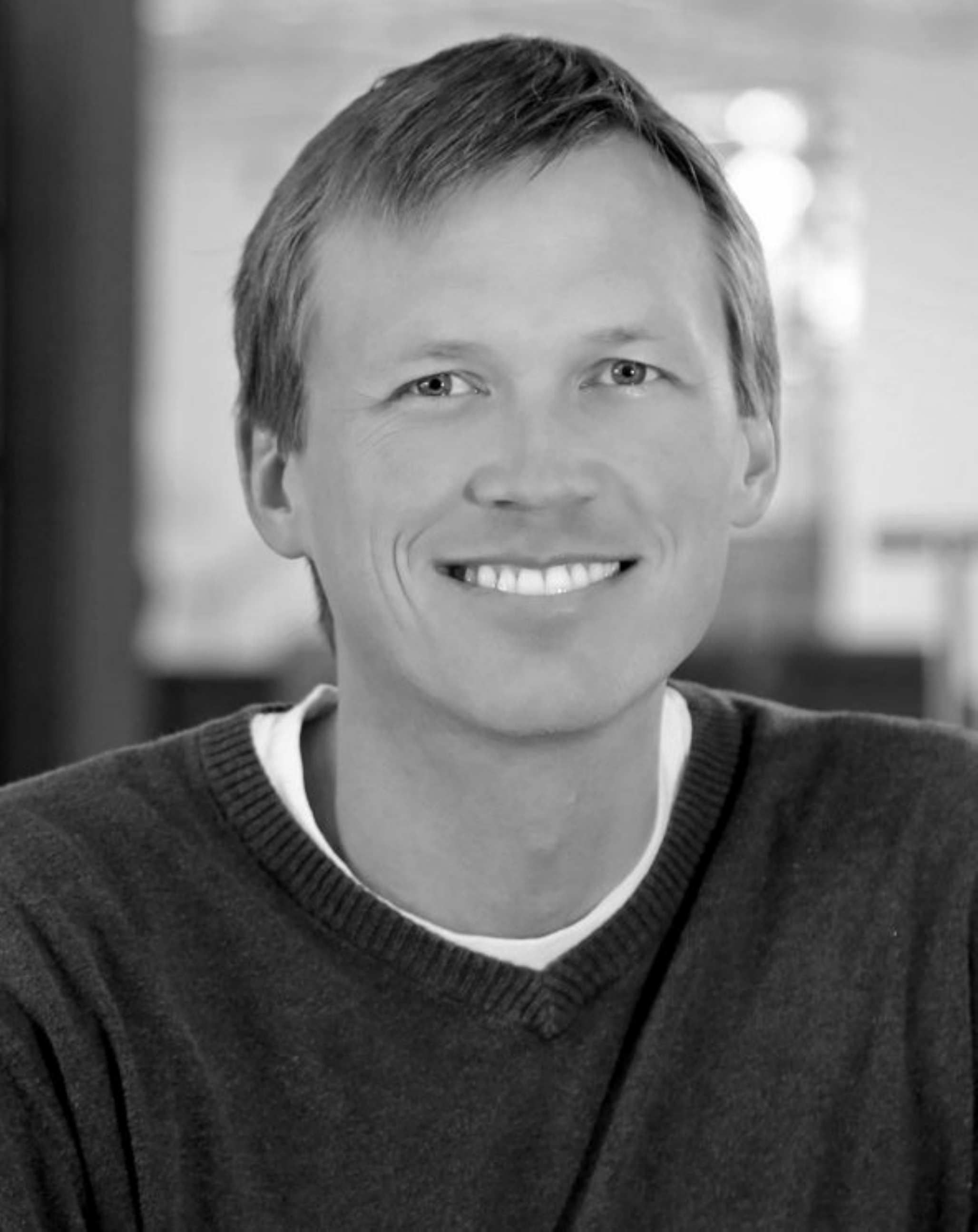}}]{Paul D.~H.~Hines} (S`96,M`07,SM`14) received the Ph.D. in Engineering and Public Policy from Carnegie Mellon University in 2007 and M.S. (2001) and B.S. (1997) degrees in Electrical Engineering from the University of Washington and Seattle Pacific University, respectively. He is currently an Associate Professor in the Dept.~of Electrical and Biomedical Engineering, with a secondary appointment in the Dept.~of Computer Science, at the University of Vermont. He also serves as the vice-chair of the IEEE PES Working Group on Cascading Failures and is a co-founder of Packetized Energy, a distributed energy software company.
\end{IEEEbiography}


\end{document}